\documentclass[12pt]{article}
\def\be{\begin{equation}}
\def\ee{\end{equation}}
\def\bea{\begin{eqnarray}}
\def\eea{\end{eqnarray}}
\usepackage{graphicx}% Include figure files

\catcode`\@=11
\def\lsim{\mathrel{\mathpalette\@versim<}}
\def\gsim{\mathrel{\mathpalette\@versim>}}
\def\@versim#1#2{\vcenter{\offinterlineskip
\ialign{$\m@th#1\hfil##\hfil$\crcr#2\crcr\sim\crcr } }}
\catcode`\@=12
\usepackage{axodraw}

\catcode`@=12
\begin{document}
\thispagestyle{empty}
\begin{flushright}
UCRHEP-T521\\
UMISS-HEP-2012-04 \\
June 2012\\
\end{flushright}
\vspace{0.5in}
\begin{center}
{\LARGE \bf Scotogenic $A_4$ Neutrino Model for\\ 
Nonzero $\theta_{13}$ and Large $\delta_{CP}$\\}
\vspace{1.5in}
{\bf Ernest Ma$^1$, Alexander Natale$^1$, and Ahmed Rashed$^{2,3}$\\}
\vspace{0.2in}
{\sl $^1$ Department of Physics and Astronomy, University of
California,\\
Riverside, California 92521, USA\\}
\vspace{0.1in}
{\sl $^2$  Department of Physics and Astronomy, 
University of Mississippi,\\ Oxford, Mississippi 38677, USA\\}
\vspace{0.1in}
{\sl $^3$  Department  of Physics, Faculty of Science, Ain Shams University,\\ 
  Cairo, 11566, Egypt\\}
\end{center}
\vspace{1.0in}
\begin{abstract}\
Assuming that neutrinos acquire radiative seesaw Majorana masses through their 
interactions with dark matter, i.e. scotogenic from the Greek 'scotos' 
meaning darkness, and using the non-Abelian discrete symmetry $A_4$, 
we propose a model of neutrino masses and mixing with nonzero $\theta_{13}$ 
and necessarily large leptonic $CP$ violation, allowing both the normal and 
inverted hierarchies of neutrino masses, as well as quasi-degenerate 
solutions.
\end{abstract}

\newpage
\baselineskip 24pt

In 2006, a one-loop mechanism was introduced~\cite{m06} linking neutrino 
mass with dark matter.  The idea is very simple.  The standard model of 
particle interactions is extended to include a second scalar doublet  
$(\eta^+, \eta^-)$ which is odd under an exactly conserved $Z_2$ 
symmetry~\cite{dm78}, as well as three neutral fermion singlets $N_i$ 
which are also odd under $Z_2$.  This requirement immediately allows 
the possibility of having the real (or imaginary) part of $\eta^0$ as a 
dark-matter candidate, which was first pointed out also in Ref.~\cite{m06}.  
As shown in Fig.~1, this results in the radiative generation of 
seesaw Majorana neutrino masses from dark matter, i.e. scotogenic from 
the Greek 'scotos' meaning darkness.
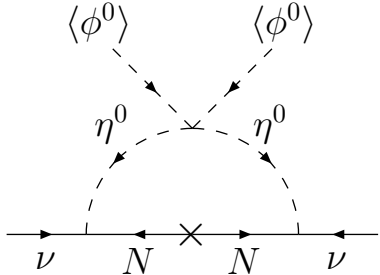
\begin{figure}[htb]
\begin{center}
\begin{picture}(260,120)(0,0)
\ArrowLine(60,10)(90,10)
\ArrowLine(130,10)(90,10)
\ArrowLine(130,10)(170,10)
\ArrowLine(200,10)(170,10)
\DashArrowArc(130,10)(40,90,180)5
\DashArrowArcn(130,10)(40,90,0)5
\DashArrowLine(100,80)(130,50)5
\DashArrowLine(160,80)(130,50)5
\Text(75,0)[]{\large $\nu$}
\Text(185,0)[]{\large $\nu$}
\Text(110,0)[]{\large $N$}
\Text(150,0)[]{\large $N$}
\Text(100,52)[]{\large $\eta^0$}
\Text(160,52)[]{\large $\eta^0$}
\Text(95,90)[]{\large $\langle \phi^0 \rangle$}
\Text(165,90)[]{\large $\langle \phi^0 \rangle$}
\Text(130,10)[]{\Large $\times$}

\end{picture}
\end{center}
\caption{One-loop generation of scotogenic Majorana neutrino mass.}
\end{figure}

The non-Abelian discrete symmetry $A_4$ was introduced~\cite{mr01,m02,bmv03} 
to achieve the seemingly impossible, i.e. the existence of a lepton family 
symmetry consistent with the three very different charged-lepton masses 
$m_e$, $m_\mu$, $m_\tau$.  It was subsequently shown~\cite{m04} to be a 
natural theoretical framework for neutrino tribimaximal mixing, i.e. 
$\sin^2 \theta_{23} = 1$, $\tan^2 \theta_{12} = 0.5$, and $\theta_{13}=0$. 
This pattern was consistent with experimental data until recently, when 
the Daya Bay Collaboration reported~\cite{daya12} the first precise 
measurement of $\theta_{13}$, i.e.
\begin{equation}
\sin^2 2 \theta_{13} = 0.092 \pm 0.016({\rm stat}) \pm 0.005({\rm syst}),
\end{equation}
followed shortly~\cite{reno12} by the RENO Collaboration, i.e.
\begin{equation}
\sin^2 2 \theta_{13} = 0.113 \pm 0.013({\rm stat}) \pm 0.019({\rm syst}).
\end{equation}
This means that tribimaximal mixing is not a good description, and more 
importantly, leptonic $CP$ violation is now possible because $\theta_{13} 
\neq 0$, just as hadronic $CP$ violation in the quark sector is possible 
because $V_{ub} \neq 0$.

Recently, it was shown~\cite{im12} that $A_4$ is still a good symmetry for 
understanding this pattern, using a new simple variation of the original 
idea~\cite{m04}.  In that proposal, neutrinos acquire Majorana masses 
through their direct interactions with Higgs triplets.  We study here 
instead the corresponding scenario with the radiative mechanism of Fig.~1.

The symmetry $A_4$ is that of 
the even permutation of four objects.  It has twelve elements and is the 
smallest group which admits an irreducible three-dimensional representation. 
Its character table is given below.
The basic multiplication rule of $A_4$ is
\begin{equation}
\underline{3} \times \underline{3} = \underline{1} + \underline{1}' + 
\underline{1}'' + \underline{3} + \underline{3}.
\end{equation}
\begin{table}[htb]
\begin{center}
\begin{tabular}{|c|c|c|c|c|c|c|}
\hline
class & $n$ & $h$ & $\chi_1$ & $\chi_{1'}$ & $\chi_{1''}$ & $\chi_3$ \\
\hline
$C_1$ & 1 & 1 & 1 & 1 & 1 & 3 \\
$C_2$ & 4 & 3 & 1 & $\omega$ & $\omega^2$ & 0 \\
$C_3$ & 4 & 3 & 1 & $\omega^2$ & $\omega$ & 0 \\
$C_4$ & 3 & 2 & 1 & 0 & 0 & --1 \\
\hline
\end{tabular}
\caption{Character table of $A_4$.}
\end{center}
\label{table-1}
\end{table}
As first shown in Ref.~\cite{mr01}, for $(\nu_i,l_i) \sim \underline{3}$, 
$l^c_i \sim \underline{1}, \underline{1}', \underline{1}''$, and 
$\Phi_i = (\phi^0_i, \phi^-_i) \sim \underline{3}$, the charged-lepton 
mass matrix is given by
\begin{equation}
{\cal M}_l = \pmatrix{v_1 & 0 & 0 \cr 0 & v_2 & 0 \cr 0 & 0 & v_3} 
\pmatrix{1 & 1 & 1 \cr 1 & \omega^2 & \omega \cr 1 & \omega & \omega^2} 
\pmatrix{f_1 & 0 & 0 \cr 0 & f_2 & 0 \cr 0 & 0 & f_3},
\end{equation}
where $v_i = \langle \phi_i^0 \rangle$ and $\omega = e^{2 \pi i/3} = -1/2 + 
i \sqrt{3}/2$.  For $v_1=v_2=v_3=v/\sqrt{3}$, 
we then obtain
\begin{equation}
{\cal M}_l = {1 \over \sqrt{3}} \pmatrix{1 & 1 & 1 \cr 1 & \omega^2 & \omega 
\cr 1 & \omega & \omega^2} \pmatrix{m_e & 0 & 0 \cr 0 & m_\mu & 0 \cr 
0 & 0 & m_\tau},
\end{equation}
where $m_e = f_1 v$, $m_\mu = f_2 v$, $m_\tau = f_3 v$. 
The original $A_4$ symmetry is now broken to the residual symmetry $Z_3$, 
i.e. lepton flavor triality~\cite{m10}, with $e \sim 1$, $\mu \sim \omega^2$, 
$\tau \sim \omega$.  This is a good symmetry of the Lagrangian as long as 
neutrino masses are zero.  Exotic scalar decays are predicted and may be 
observable at the Large Hadron Collider (LHC) in some regions of parameter 
space~\cite{ckmo11,cdmw11}.

To obtain nonzero neutrino masses, we assign $\eta \sim \underline{1}$ and 
$N_i \sim \underline{3}$ under $A_4$.  We also add the scalar singlets 
$\sigma_i \sim \underline{3}$ with nonzero $\langle \sigma_i \rangle$.  
The resulting $3 \times 3$ Majorana mass matrix for $N_i$ is then
\begin{equation}
{\cal M}_N = \pmatrix{A & F & E \cr F & A & D \cr E & D & A},
\end{equation}
which is the analog of
\begin{equation}
{\cal M}_\nu = \pmatrix{a & f & e \cr f & a & d \cr e & d & a},
\end{equation}
considered in Ref.~\cite{im12}.  (A better way to enforce Eq.~(6) is to 
postulate gauged $B-L$ and assume complex neutral scalars which transform as 
$\underline{1},~\underline{3}$ under $A_4$, in complete analogy to the 
scalar triplets of Ref.~\cite{im12}.)  Instead of enforcing $E=F=0$ which is 
required for tribimaximal mixing, we assume here that $F=-E$ which may 
be maintained by an interchange symmetry~\cite{m04,mw11}.

Consider now the tribimaximal basis, i.e.
\begin{equation}
\pmatrix{\nu_e \cr \nu_\mu \cr \nu_\tau} = \pmatrix{\sqrt{2/3} & 1/\sqrt{3} 
& 0 \cr -1/\sqrt{6} & 1/\sqrt{3} & -1/\sqrt{2} \cr -1/\sqrt{6} & 1/\sqrt{3} 
& 1/\sqrt{2}} \pmatrix{\nu_1 \cr \nu_2 \cr \nu_3}.
\end{equation}
Since $\nu_{1,2,3}$ are connected to $N_{1,2,3}$ through the identity matrix, 
we find
\begin{equation}
{\cal M}_N^{(1,2,3)} = \pmatrix{A+D & 0 & 0 \cr 0 & A & C \cr 0 & C & A-D},
\end{equation}
where $C=(E-F)/\sqrt{2}=\sqrt{2}E$.

The diagram of Fig.~1 is exactly calculable from the exchange of Re($\eta^0$) 
and Im($\eta^0$) and is given by~\cite{m06}
\begin{equation}
({\cal M}_\nu)_{ij} = \sum_k {h_{ik} h_{jk} M_k \over 16 \pi^2} \left[ 
{m_R^2 \over m_R^2 - M_k^2} \ln {m_R^2 \over M_k^2} - 
{m_I^2 \over m_I^2 - M_k^2} \ln {m_I^2 \over M_k^2} \right],
\end{equation}
where $\sum_k h_{ik} (h_{jk})^* = |h|^2 \delta_{ij}$, and $m_{R,I}$ are the masses 
of $\sqrt{2}$Re($\eta^0$) and $\sqrt{2}$Im($\eta^0$), respectively.  In the 
limit $m_R^2-m_I^2 = 2 \lambda_5 v^2$ is small compared to $m_0^2 = 
(m_R^2 + m_I^2)/2$, and $m_0^2 << M_k^2$, Eq.~(10) reduces to
\begin{equation}
({\cal M}_\nu)_{ij} = {\lambda_5 v^2 \over 8 \pi^2} \sum_k {h_{ik} h_{jk} 
\over M_k} \left[ \ln {M_k^2 \over m_0^2} - 1 \right].
\end{equation}
In the tribimaximal basis of Eq.~(9), we then have
\begin{equation}
h_{ik} = h \pmatrix{1 & 0 & 0 \cr 0 & \cos \theta & -\sin \theta e^{i \phi}  
\cr 0 & \sin \theta e^{-i \phi} & \cos \theta} \pmatrix{e^{i \alpha'_1/2} 
& 0 & 0 \cr 0 & e^{i \alpha'_2/2} & 0 \cr 0 & 0 & e^{i \alpha'_3/2}},
\end{equation}
with
\begin{equation}
\pmatrix{\cos \theta & \sin \theta e^{i \phi} \cr 
-\sin \theta e^{-i \phi} & \cos \theta} \pmatrix{A & C \cr C & A-D} 
\pmatrix{\cos \theta & -\sin \theta e^{-i \phi} 
\cr \sin \theta e^{i \phi} & \cos \theta} = \pmatrix{e^{i \alpha'_2} M_2 
& 0 \cr 0 & e^{i \alpha'_3} M_3}.
\end{equation}

The neutrino mixing matrix $U$ has 4 parameters: $s_{12}, s_{23}, s_{13}$ and
$\delta_{CP}$~\cite{pdg10}.  We choose the convention $U_{\tau 1}, U_{\tau 2},
U_{e3}, U_{\mu 3} \to -U_{\tau 1}, -U_{\tau 2},
-U_{e3}, -U_{\mu 3}$ to conform with that of the tribimaximal mixing matrix
of Eq.~(8), then
\begin{equation}
{\cal M}_\nu^{(1,2,3)} = U^T_{TB} U \pmatrix{e^{i\alpha_1} m'_1 & 0 & 0 \cr
0 & e^{i\alpha_2} m'_2 & 0 \cr 0 & 0 &  m'_3} U^T U_{TB},
\end{equation}
where $m'_{1,2,3}$ are the physical neutrino masses, with
\begin{eqnarray}
m'_2 &=& \sqrt{{m'_1}^2 + \Delta m^2_{21}}, \\
m'_3 &=& \sqrt{{m'_1}^2 + \Delta m^2_{21}/2 + \Delta m_{32}^2}~~{\rm (normal
~hierarchy)}, \\
m'_3 &=& \sqrt{{m'_1}^2 + \Delta m^2_{21}/2 - \Delta m_{32}^2}~~{\rm
(inverted~hierarchy)}.
\end{eqnarray}
We now diagonalize ${\cal M}_\nu^{(1,2,3)}$ using
\begin{equation}
U_\epsilon {\cal M}_\nu^{(1,2,3)} U_\epsilon^T = \pmatrix{e^{i \alpha'_1}
m'_1 & 0 & 0 \cr 0 & e^{i \alpha'_2} m'_2 & 0 \cr 0 & 0 & e^{i \alpha'_3} 
m'_3},
\end{equation}
from which we obtain $U' = U_{TB} U^T_\epsilon$.  To obtain $U$ with the 
usual convention, we rotate the phases of the $\mu$ and $\tau$ rows so that
$U'_{\mu 3} e^{-i \alpha'_3/2}$ is real and negative, and 
$U'_{\tau 3} e^{-i \alpha'_3/2}$ is real and positive.
These phases are absorbed by the $\mu$ and $\tau$ leptons and are
unobservable.  We then rotate the $\nu_{1,2}$ columns so that 
$U'_{e1} e^{-i \alpha'_3/2} = U_{e1} e^{i \alpha''_1/2}$ and 
$U'_{e2} e^{-i \alpha'_3/2} = U_{e2} e^{i \alpha''_2/2}$, where 
$U_{e1}$ and $U_{e2}$ are real and positive.  
The physical relative Majorana phases of $\nu_{1,2}$ are then 
$\alpha_{1,2} = \alpha'_{1,2} + \alpha''_{1,2}$.  The three angles and 
the Dirac phase are extracted according to
\begin{equation}
\tan^2 \theta_{12} = |U'_{e2}/U'_{e1}|^2, ~~~ 
\tan^2 \theta_{23} = |U'_{\mu 3}/U'_{\tau 3}|^2, ~~~ 
\sin \theta_{13} e^{-i\delta_{CP}} = U'_{e3} e^{-i \alpha'_3/2}.
\end{equation}
The effective Majorana neutrino mass in neutrinoless double beta decay is 
then given by
\begin{equation}
m_{ee} = |U_{e1}^2 e^{i \alpha_1} m'_1 + U_{e2}^2 e^{i \alpha_2}
m'_2 + U_{e3}^2 m'_3|.
\end{equation}

In Eq.~(9), let $A$ be real and positive by convention, then both $C$ and 
$D$ may be complex, i.e. $C = C_R + iC_I$ and $D = D_R + iD_I$.  The 
$2 \times 2$ matrix of Eq.~(13) can be solved exactly to yield
\begin{eqnarray}
\tan \phi &=& {C_R D_I - C_I D_R \over C_R(2A-D_R) - C_I D_I}, \\ 
\tan 2 \theta &=& {2 [4 A^2 C_R^2 - 4 A C_R (C_R D_R + C_I D_I) + 
(C_R^2 + C_I^2) (D_R^2 + D_I^2)]^{1/2} \over 2A D_R - (D_R^2 + D_I^2)},
\end{eqnarray}
with
\begin{eqnarray}
e^{i \alpha'_2} M_2 &=& \cos^2 \theta A + 2 \sin \theta \cos \theta e^{i \phi} 
C + \sin^2 \theta e^{2 i \phi} (A-D), \\
e^{i \alpha'_3} M_3 &=& \cos^2 \theta (A-D) - 2 \sin \theta \cos \theta 
e^{-i \phi} C + \sin^2 \theta e^{-2 i \phi} A.
\end{eqnarray}
The corresponding $U'$ elements are
\begin{eqnarray}
&& U'_{e1} = \sqrt {2 \over 3}, ~~~ U'_{e2} = {\cos \theta \over \sqrt{3}}, 
~~~ U'_{e3} = - {\sin \theta \over \sqrt{3}} e^{-i \phi}, \\ 
&& U'_{\mu 3} = - {\cos \theta \over \sqrt{2}} - {\sin \theta \over \sqrt{3}} 
e^{-i \phi}, ~~~  U'_{\tau 3} = {\cos \theta \over \sqrt{2}} - {\sin \theta 
\over \sqrt{3}} e^{-i \phi}.
\end{eqnarray}

If we absorb the scale factor $\lambda_5 h^2 v^2/8 \pi^2$ into the parameters 
$A,C,D$ as well as $m_0$, then the mass eigenvalues of Eq.~(11) are given by
\begin{equation}
m'_k = {1 \over M_k} \left[ \ln {M_k^2 \over m_0^2} - 1 \right], 
\end{equation}
which are the ones used in Eqs.~(14) and (18).  Since $m_0$ is an unknown, 
having to do with the dark-matter scalar mass, we fix it by requiring 
$M_1 / m_0 = 10$, where $M_1 = |A+D|$.  If we input the five parameters 
$A,C_R,C_I,D_R,D_I$, we will obtain $m'_{1,2,3}$ as well as the three 
mixing angles and the three $CP$ phases.  For our numerical analysis, we 
set
\begin{equation}
\Delta m^2_{21} = 7.59 \times 10^{-5}~{\rm eV}^2, ~~~ 
\Delta m^2_{32} = 2.45 \times 10^{-3}~{\rm eV}^2, 
\end{equation}
and vary $\theta_{13}$ in the range
\begin{equation}
\sin^2 2 \theta_{13} = 0.05~{\rm to}~0.15.
\end{equation}
Following Ref.~\cite{im12}, we look for solutions with $\sin^2 2 \theta_{23} 
= 0.92$ and 0.96.  Whereas only normal hierarchy is allowed in the model 
of Ref.~\cite{im12}, we find solutions for both normal and inverted 
hierarchies, as well as quasi-degenerate solutions, as detailed below.

The predictions of this model regarding mixing angles are basically the 
same as in Ref.~\cite{im12} for the special case of $b=0$ there.  
Using Eqs.~(19), (25), and (26), we find
\begin{eqnarray}
\tan^2 \theta_{12} &=& {1 - 3 \sin^2 \theta_{13} \over 2}, \\ 
\tan^2 \theta_{23} &=& { \left(1- {\sqrt{2} \sin \theta_{13} \cos \phi 
\over \sqrt{1-3 \sin^2 \theta_{13}}} \right)^2 + {2 \sin^2 \theta_{13} 
\sin^2 \phi \over 1-3 \sin^2 \theta_{13}} \over 
\left(1+ {\sqrt{2} \sin \theta_{13} \cos \phi 
\over \sqrt{1-3 \sin^2 \theta_{13}}} \right)^2 + {2 \sin^2 \theta_{13} 
\sin^2 \phi \over 1-3 \sin^2 \theta_{13}}}.
\end{eqnarray}
The conventionally defined Dirac $CP$ phase is given by 
$\delta_{CP} = \phi + \alpha'_3/2$, where $\alpha'_3$ 
is defined in Eq.~(18) and depends on the specific values of Eq.~(9). 
For $\sin \theta_{13} = 0.16$, corresponding to $\sin^2 2 \theta_{13} = 0.1$, 
this predicts $\tan^2 \theta_{12} = 0.46$.   If $Im(C)=0$, then 
$\delta_{CP}=\alpha'_3 = 0$, so this 
would predict $\sin^2 2 \theta_{23} = 0.80$ which is of course ruled 
out.  Using $\sin^2 2 \theta_{23} > 0.92$, we find in this case 
$|\tan \phi| > 1.2$.

\begin{figure}[htb]
	\centering
		\includegraphics[width=8cm]{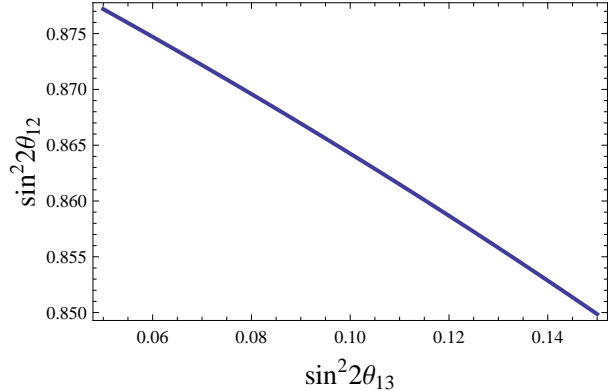}
	\caption{$\sin^2 2 \theta_{12}$ versus $\sin^2 2 \theta_{13}$.}
	\label{sines}
\end{figure}

For each of the two values $\sin^2 2 \theta _{23} = 0.92$ and 0.96, we 
obtain 5 representative solutions, all as functions of $\sin^2 2 \theta_{13}$. 
Using Eq.~(30), we plot $\sin^2 2 \theta_{12}$ versus $\sin^2 2 \theta_{13}$ 
in Fig.~\ref{sines}.  
The characteristic features of the 5 solutions are listed 
in Table~2.
\begin{table}[htb]
\begin{center}
\begin{tabular}{|c|c|c|c|c|}
\hline
solution & $Im(D)$ & class & $|\tan \delta_{CP}|$ & $m_{ee}$ \\
\hline
I & 0 & IH & 2.05 & 0.020 \\
II & $Re(D)$ & IH & 4.64 & 0.022 \\
III & 0 & NH & 3.59 & 0.002 \\
IV & 0 & QD & 2.20 & 0.046 \\
V & $Re(D)$ & QD & 1.84 & 0.051 \\
\hline
\end{tabular}
\caption{Five representative solutions. 
Three have $Im(D)=0$, and two have $Im(D)=Re(D)$.  NH denotes normal 
hierarchy of neutrino masses, IH inverted, and QD quasi-degenerate. 
The values of $|\tan \delta_{CP}|$ and $m_{ee}$ (in eV) are for 
$\sin^2 2 \theta_{23} = 0.96$ and $\sin^2 2 \theta_{13}=0.10$.}
\end{center}
\label{table2}
\end{table}
For $Im(D)=0$, we find one solution for inverted ordering of neutrino masses, 
and two solutions for normal ordering (one of which is quasi-degenerate). 
For $Im(D)=Re(D)$, we again find one solution for inverted ordering, but 
the only solution for normal ordering is quasi-degenerate.

In Fig.~3 we show the physical neutrino masses $m'_{1,2,3}$ and the effective 
mass in neutrinoless double beta decay $m_{ee}$ (in eV) as well as the model 
parameters (in eV$^{-1}$) for solution (I) in the case 
$\sin^2 2 \theta_{23} = 0.96$.  In Figs.~4-7 we show the same quantities 
for solutions (II),(III),(IV),(V) in the cases of $\sin^2 2 \theta_{23} 
= 0.92, 0.96, 0.92, 0.96$ respectively.  Finally we show in Fig.~8 
the values of $|\tan \delta_{CP}|$ for all 5 solutions in the case of 
$\sin^2 2 \theta_{23} = 0.92$.  It is claer that at 
$\sin^2 2 \theta_{13} = 0.10$, large $|\tan \delta_{CP}|$ is predicted.

%\clearpage

%\newpage
\underline{Acknowledgments}: 
This work is supported in part by the U.~S.~Department of Energy 
under Grant No.~DE-AC02-06CH11357. The work of A.R. is supported 
in part by the National Science Foundation under Grant No.\ NSF PHY-1068052 and in part by the Graduate Student Council Research Grant Award 2012-2013. A.R. thanks the University of California, Riverside for hospitality.

%\newpage
\baselineskip 16pt
\bibliographystyle{unsrt}

\clearpage

\begin{figure}[htb]
	\centering
		\includegraphics[width=8cm]{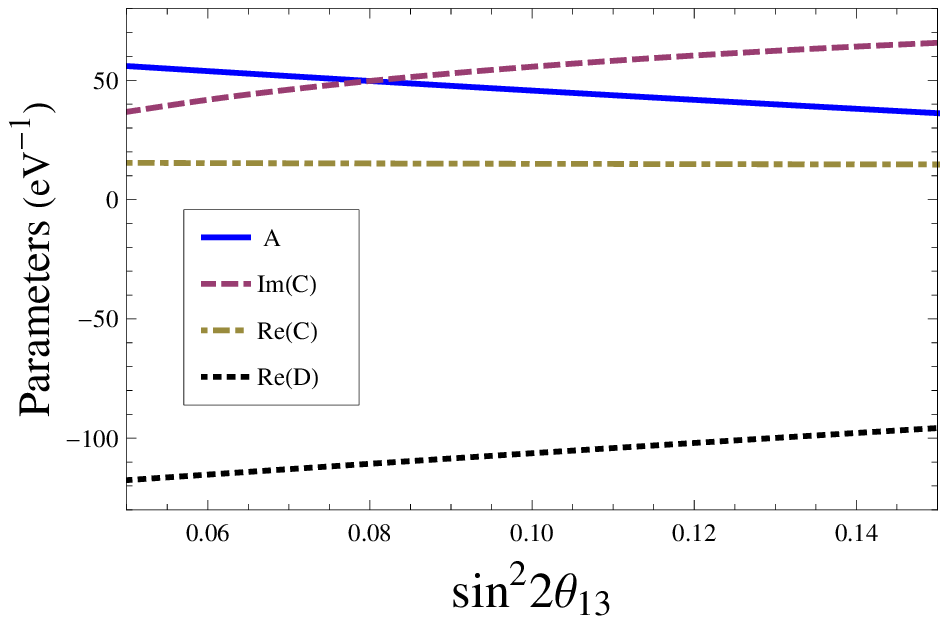}
		\includegraphics[width=8cm]{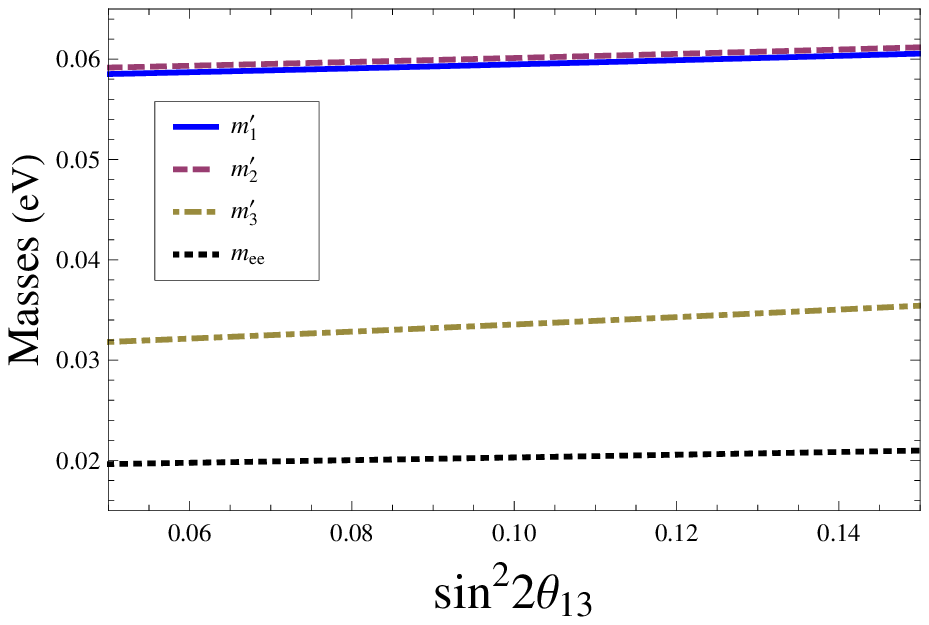}
	\caption{$A_4$ parameters and the physical neutrino masses and effective neutrino mass $m_{ee}$ in neutrinoless double beta decay for the inverted hierarchy with Im(D)=0 and $\sin^22\theta_{23}=0.96$.}
	\label{plot-1}
\end{figure}

\begin{figure}[htb]
	\centering
		\includegraphics[width=8cm]{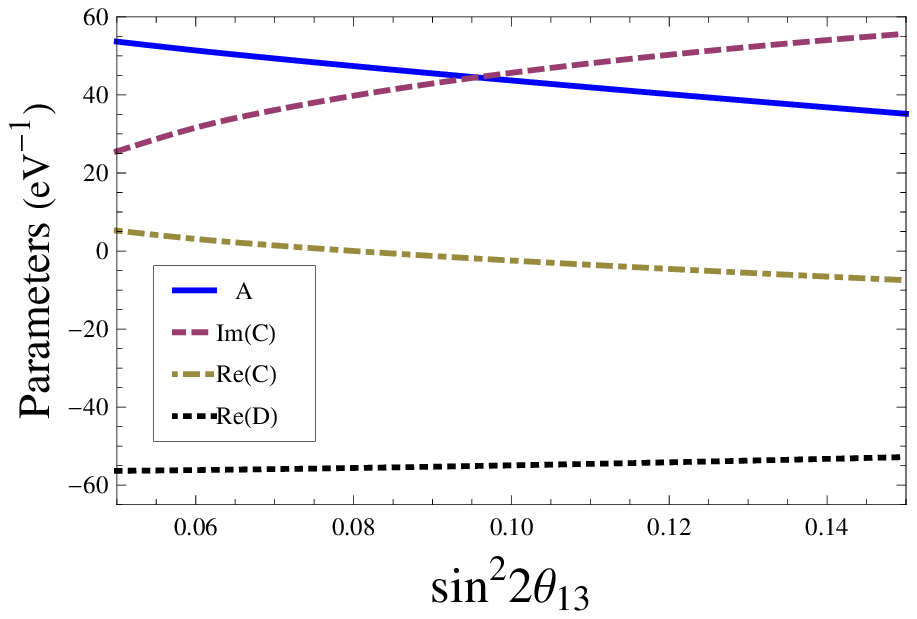}
		\includegraphics[width=8cm]{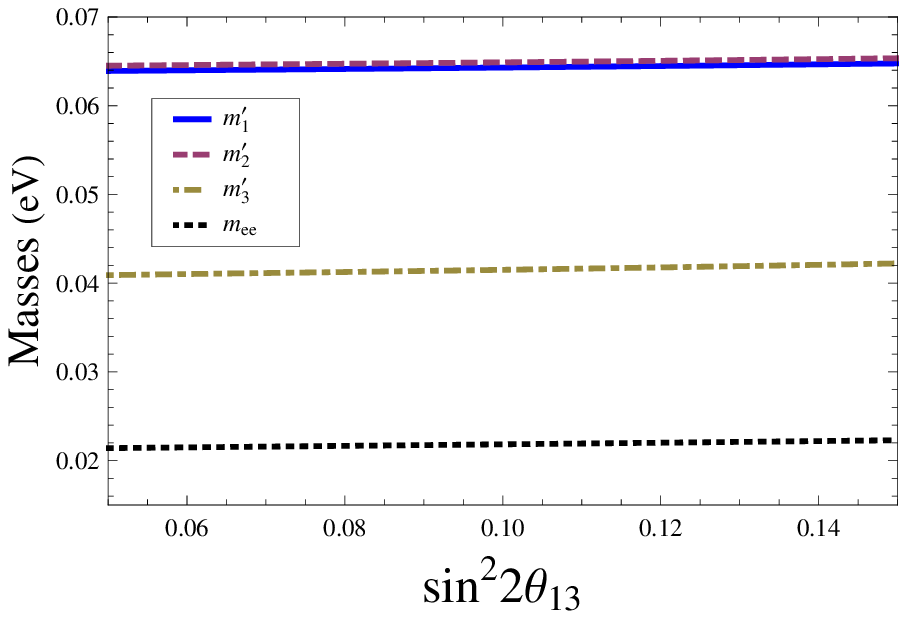}
	\caption{$A_4$ parameters and the physical neutrino masses and effective neutrino mass $m_{ee}$ in neutrinoless double beta decay for the inverted hierarchy with Im(D)=Re(D) and $\sin^22\theta_{23}=0.92$.}
	\label{plot-2}
\end{figure}

\clearpage

\begin{figure}[htb]
	\centering
		\includegraphics[width=8cm]{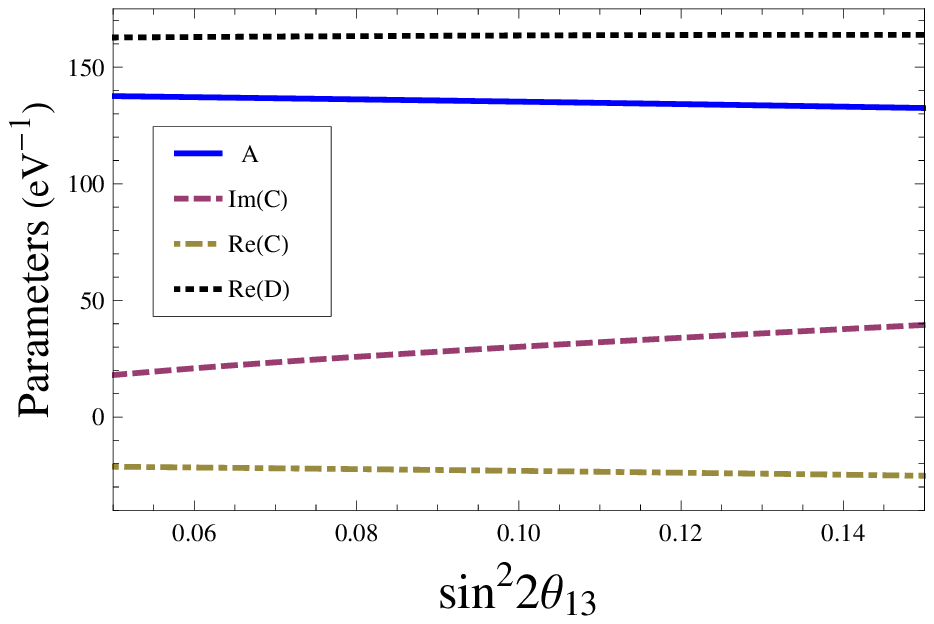}
		\includegraphics[width=8cm]{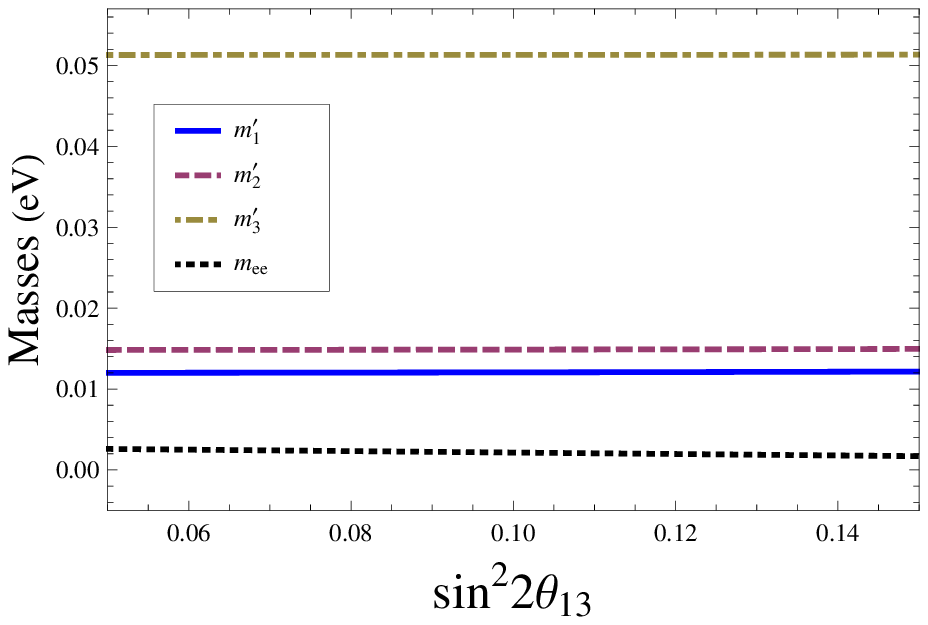}
	\caption{$A_4$ parameters and the physical neutrino masses and effective neutrino mass $m_{ee}$ in neutrinoless double beta decay for the normal hierarchy with Im(D)=0 and $\sin^22\theta_{23}=0.96$.}
	\label{plot-3}
\end{figure}

\begin{figure}[htb]
	\centering
		\includegraphics[width=8cm]{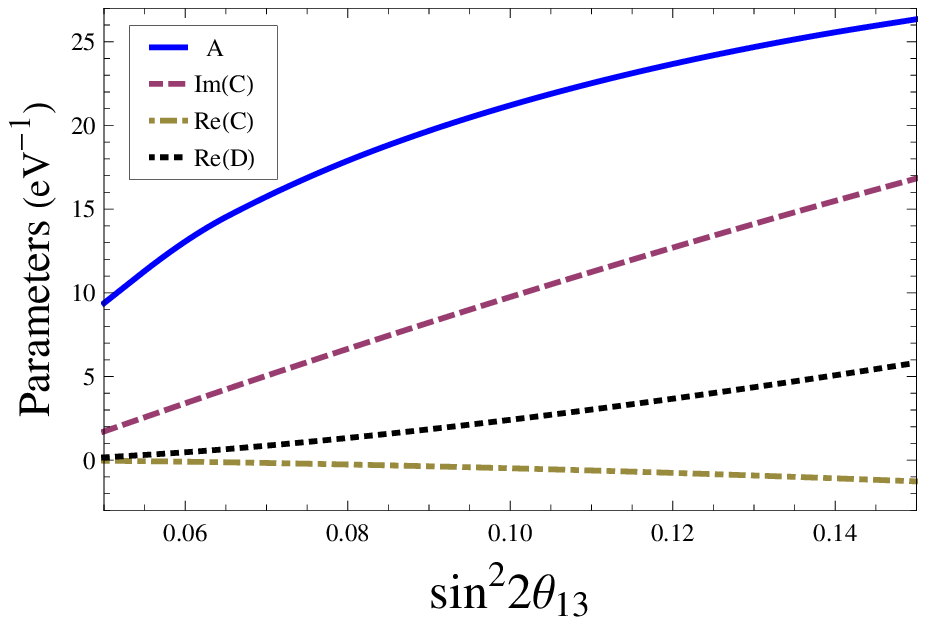}
		\includegraphics[width=8cm]{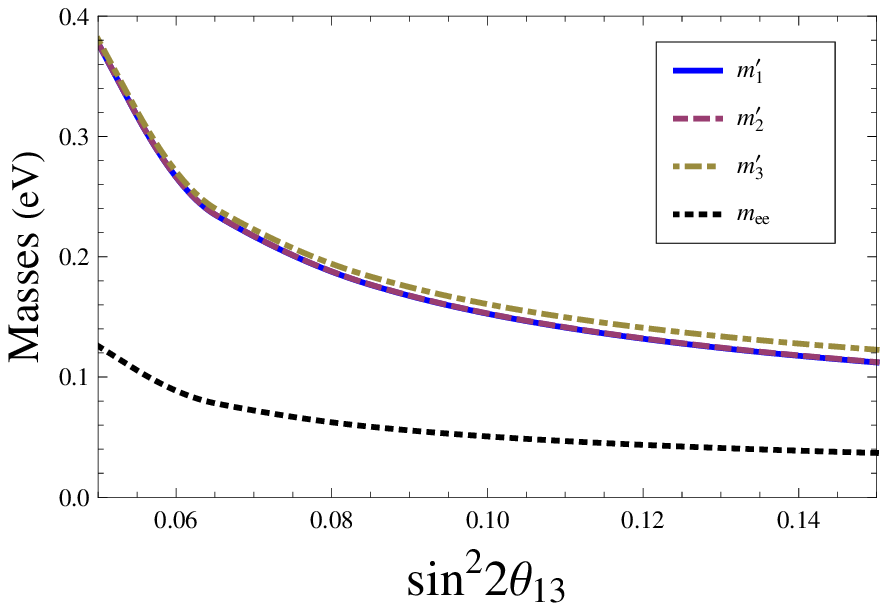}
	\caption{$A_4$ parameters and the physical neutrino masses and effective neutrino mass $m_{ee}$ in neutrinoless double beta decay for quasi-degenerate neutrino masses with Im(D)=0 and $\sin^22\theta_{23}=0.96$.}
	\label{plot-4}
\end{figure}

%\clearpage

\begin{figure}[htb]
	\centering
		\includegraphics[width=8.0cm]{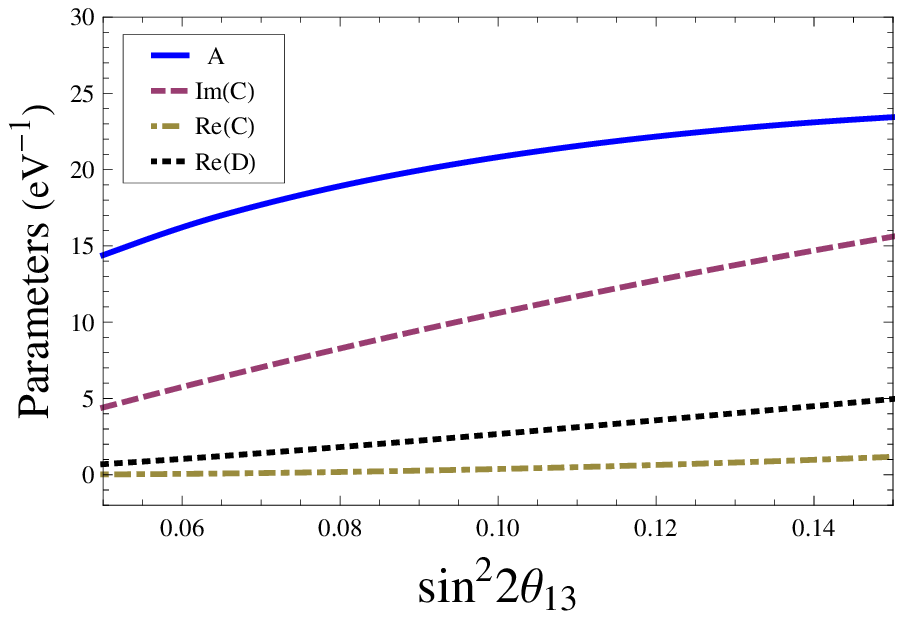}
		\includegraphics[width=8.0cm]{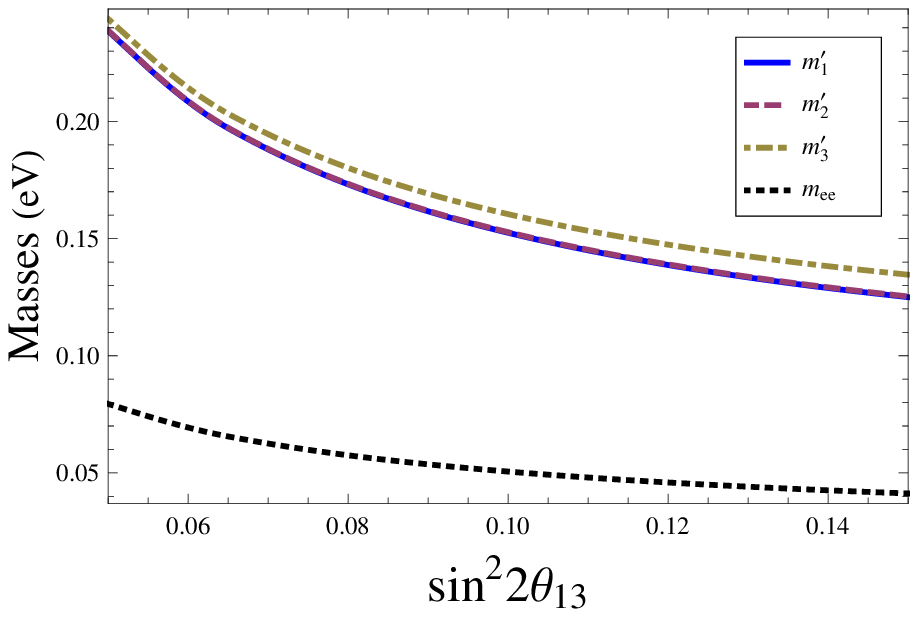}
	\caption{$A_4$ parameters and the physical neutrino masses and effective neutrino mass $m_{ee}$ in neutrinoless double beta decay for quasi-degenerate neutrino masses with Im(D)=Re(D) and $\sin^22\theta_{23}=0.96$.}
	\label{plot-5}
\end{figure}

\begin{figure}[htb]
	\centering
		\includegraphics[width=14cm]{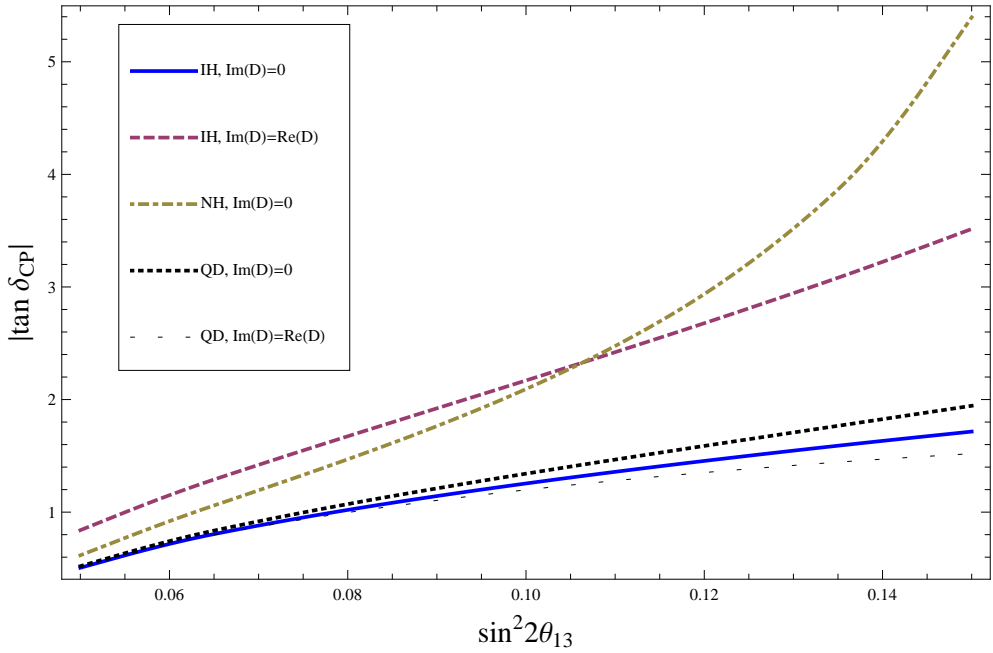}
	\caption{$|\tan \delta_{CP}|$ versus $\sin^2 2 \theta_{13}$ for $\sin^22\theta_{23}=0.92$.}
	\label{plot-6}
\end{figure}

\end{document}